\documentstyle[aps,prl,floats,epsfig]{revtex}

\begin{document}


\title{A Microscopic Model of Ferroelectricity in Stress-free PbTiO$_3$ 
Ultrathin Films}
\author{Ph. Ghosez\cite{a)} 
and K. M. Rabe}
\address{Department of Applied Physics, Yale University, 
P.O. Box 208284, New Haven, CT 06520-8284, USA.}
\date{\today}
\maketitle

\begin{abstract}
The ground-state polarization of PbTiO$_3$ thin films is studied 
using a microscopic effective Hamiltonian with parameters obtained
from first-principles calculations. Under short-circuit electrical 
boundary conditions, (001) films with thickness as low as three 
unit cells are found to have a perpendicularly polarized ferroelectric 
ground state with significant enhancement of the polarization at 
the surface.
\end{abstract}

\vskip1pc
    
Ferroelectricity is a collective phenomenon and, as such, it is 
expected to be strongly influenced by surfaces and finite size 
effects~\cite{Lines77,Scott98}. 
It has been generally thought that decreasing the thickness of thin 
films of perovskite ABO$_3$ compounds suppresses ferroelectricity 
and eliminates it altogether at a nonzero critical thickness. 
In PbTiO$_3$, this critical size has been estimated 
as 7-10 nm \cite{Ishikawa96,Li97,Maruyama98}. However, recently 
synthesized high quality films of PbZr$_{x}$Ti$_{1-x}$O$_{3}$ have 
been observed to exhibit a stable perpendicularly polarized 
ferroelectricity down to and possibly even below a thickness of 40 
\AA \cite{TybellU}.
This suggests that the earlier observations could stem from 
variations in electrical and mechanical boundary conditions or from 
extrinsic effects associated with defects and impurities.

In this Letter, we address the question of the intrinsic 
ferroelectricity of PbTiO$_3$ thin films using a first-principles 
effective Hamiltonian (${\cal H}_{eff}$) based on that previously 
developed for studies of the finite-temperature ferroelectric 
transition in the bulk~\cite{Waghmare97}. 
This approach combines the advantages of phenomenological models, 
which are physically transparent, with those of 
first-principles simulations, which yield highly accurate 
material-specific microscopic information. 
We find that PbTiO$_3$ films exhibit a ferroelectric instability 
which is even greater than that of the bulk, and identify the 
reason for this within the model. Further, we analyze the spatial 
variation of the polarization within the film, showing that 
polarization is significantly enhanced over the bulk value in a 
single surface layer.

We consider symmetrically TiO$_2$ terminated (001) slabs of PbTiO$_3$, 
under stress-free and short-circuit boundary conditions. The latter 
are imposed by the presence of perfectly conducting sheets positioned at 
a distance a$_0$/4 above and below the top and bottom surfaces of the 
slab, where a$_0$ is the experimental lattice constant of bulk PbTiO$_3$ 
(a$_0$=3.97 \AA). 

As for the bulk \cite{Waghmare97}, for the thin films, ${\cal H}_{eff}$ 
is constructed by projecting the full interionic Hamiltonian 
into a subspace which contains the relevant degrees of freedom. These 
are the localized atomic displacement patterns ($\vec{\xi_{i}}$)
centered around each unit cell $i$, determined by the lattice 
Wannier function method \cite{LWF}, and the homogeneous strain tensor 
$e_{\alpha\beta}$. The form of the resulting model is
$${\cal H}_{eff} = ({\cal H}_{dipolar} + {\cal H}_{sr}) + 
{\cal H}_{anharm} + {\cal H}_{strain}$$ where the harmonic part
is separated into a long-range dipolar term, ${\cal H}_{dipolar}$, and 
short-range corrections, ${\cal H}_{sr}$. ${\cal H}_{anharm}$ represents
the anharmonic couplings and ${\cal H}_{strain}$ combines the elastic
energy and the lowest order coupling of $\vec{\xi_{i}}$ to $e_{\alpha\beta}$.
The local dipole per unit cell is $\vec{p_{i}}= Z^{*}_{i}ea_0 \vec{\xi_{i}}$ 
where $Z^{*}_{i}$ is the Born effective charge associated with $\vec{\xi_{i}}$. 

The present construction, based on projection from a thin-film 
interionic Hamiltonian, is substantially refined over that 
presented previously \cite{RabeU}, which was a simple slab 
truncation of the bulk ${\cal H}_{eff}$. However, in order to avoid 
computationally intensive first-principles slab calculations
\cite{surfBT,surfBTa,Fu98}, we use an approximate interionic 
Hamiltonian, obtained by transfer of microscopic information 
available for the bulk to the thin film geometry. The associated 
modifications include termination of the interatomic short-range 
force constants at the surface \cite{Ghosez99}, change in the effective 
dipole-dipole interaction resulting from the perfectly conducting 
plates which implement the short-circuit boundary conditions, and 
corrections to preserve global translational symmetry and charge 
neutrality~\cite{Ruini98}.
The projection is then performed using the same basis functions as 
for the bulk, with minor modifications for the unit cells at the 
surface. The anharmonic and strain terms are, for simplicity, 
assumed to be the same as in the bulk. Further details will be given 
in Ref. \onlinecite{inprep}.

\begin{figure}
\vspace{0.5cm}
\epsfig{file=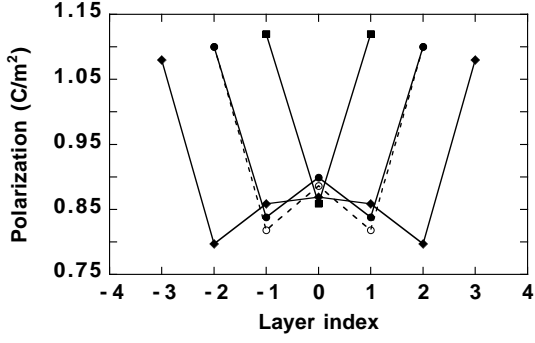,width=7cm}
\vspace{0.5cm}
\caption{
Layer-by-layer profiles of the polarization for films of 
thickness N=3 (squares), N=5 (solid circles) and N=7 (diamonds). 
The polarization profile for the simplified model for N=5 is shown by 
the open circles and dashed line. For the bulk, the  computed polarization 
in the tetragonal phase is equal to 0.83 C/m$^{2}$.}
\label{Fig.profile}
\end{figure}

The minimum of ${\cal H}_{eff}$ for in-plane and 
perpendicularly polarized films is reported for different thicknesses 
in Table~\ref{Table.energy}. 
The polarization is assumed to be uniform within each layer. The 
mechanical boundary conditions are taken as stress free, so that the 
strain relaxes to its optimum value. At each thickness,
perpendicular polarization is seen to be most favorable, with 
stabilization relative to the paraelectric state increasing with 
decreasing thickness. The corresponding layer-by-layer polarization 
profiles are shown in Fig.~\ref{Fig.profile}. 
While in the interior the polarization approaches the bulk value even 
for very thin films, it is significantly enhanced at the surface.
In fact, unpolarized films are unstable against polarization of only 
the two surface layers. 
The enhancement of the average polarization couples to the strain, 
resulting in a c/a ratio that increases slightly with decreasing thickness.

To understand the ferroelectric instability and surface polarization 
enhancement in perpendicularly polarized films, we consider 
${\cal H}_{eff}$ specialized to the case of perpendicular polarization, 
constant in each layer. The computed parameters are both layer and 
thickness dependent, approaching bulk values in the interior of the film.
As illustrated in Table I and in Figure 
\ref{Fig.profile}, 
this behavior can be closely reproduced by assigning bulk values to 
all parameters except the onsite interaction in the surface layer and 
the quadratic interaction between the surface and its neighboring 
layer, and by approximating the dipolar interaction matrix as described 
below. This corresponds to the following {\it simplified} model with
parameters reported in Table~\ref{Table.para}. The short-range and 
anharmonic part take the form:
\begin{eqnarray}
{\cal H}_{sr}+{\cal H}_{anharm} &=&
\sum_{n=1}^N (A_l \; \xi_n^2 + B \; \xi_n^4 + C \; \xi_n^6 + D \; \xi_n^8) +
\sum_{n=1}^{N-1} a \; \xi_n \; \xi_{n+1} + 
\sum_{n=1}^{N-2} b \; \xi_n \; \xi_{n+2}
\nonumber \\
&& + 
\Delta A_l \; (\xi_1^2 + \xi_N^2)+
\Delta a \; (\xi_1 \; \xi_2 + \xi_{N-1} \; \xi_N) \nonumber .
\end{eqnarray}
where $A_l, B, C, D, a $ and $b$ are directly related to 
bulk ${\cal H}_{eff}$ parameters, while $\Delta A_l$ and $\Delta a$ arise from 
changes at the surfaces and are, to a good 
approximation, independent of the thickness.

The dipolar contribution, computed using the 
Ewald summation technique including the effect of the metallic 
plates, is accurately represented (to within 0.1\%) by the expression
\begin{eqnarray}
{\cal H}_{dipolar}&=& \frac{1}{2 \epsilon_{\infty}} [ \frac{4 \pi 
u_{conv}}{a_{0} (N+1/2)}
(\sum_{n=1}^N Z^{*}_{n} \xi_n)^2 +u_{self} \sum_{n=1}^N Z^{* 2}_{n} 
\xi_n^2 \\ &&
+u_{sur} (Z^{* 2}_{1} \xi_{1}^{2}+ Z^{* 2}_{N} \xi_{N}^{2})
+u_{nn} \sum_{n=1}^{N-1} Z^{*}_{n} Z^{*}_{n+1} \xi_{n} \xi_{n+1}].
\nonumber
\end{eqnarray}
where the optical dielectric constant, $\epsilon_{\infty}$, is taken 
equal to 8.24~\cite{Waghmare97,Ghosez99} and $Z^{*}$ is $+10.218$ 
at the surface and $+10.056$ elsewhere.

Finally, the terms which represent the coupling of the polarization 
to homogeneous strain are derived directly from the corresponding 
bulk terms:
\begin{eqnarray}
{\cal H}_{strain}={N\over 2} C_{11}\sum_\alpha e_{\alpha\alpha}^2 + 
NC_{12}\sum_{\alpha <\beta}e_{\alpha\alpha} e_{\beta\beta}  +
(\sum_{n=1}^N \xi_n^2)(g_0 \sum_\alpha e_{\alpha\alpha} + g_1 
e_{33}).
\nonumber
\end{eqnarray}

Within this simplified model, the origin of the enhancement of the 
ferroelectric instability in thin films over the bulk can be readily 
identified. The model can be obtained from a reference model which has the
uniform bulk ferroelectric ground state by modifying three terms as 
follows:
(i) changing the dipolar contribution by moving the perfectly 
conducting plates, at the level of the model, 
outward by a$_0$/4 from where the array of dipoles combined with their
images forms an infinite 
simple cubic lattice;
(ii) suppressing the periodic boundary conditions on the short-range
terms, that is, eliminating the coupling of layers 1 and 2 to layers N and
N+1 , 
and (iii) including the short-range surface corrections $\Delta A_l$ 
and $\Delta a$.
We can introduce these modifications of the bulk model separately.

By examining first the effects of term (i), 
we can qualitatively understand how the electrical boundary condition 
plays a crucial role in determining the thin film ground
state~\cite{plates}.
The change in the dipolar contribution from the bulk to the thin film 
significantly suppresses the perpendicular polarization. This 
is not surprising as motion of
the plates outward from the slab decreases the compensation of the 
depolarization field.  In contrast, starting from the bulk and 
truncating the antiferroelectric short-range couplings
at the surface (term (ii)) leads to a large energy gain, a large 
enhancement of the surface polarization, and a somewhat lesser 
enhancement of the polarization of the interior layers which is close 
to constant starting with the second layer. Examining the effects of
the 
short-range surface terms $\Delta A_l$ and $\Delta a$ 
(term (iii)), we find similar behavior, except that the polarization 
in the second layer, while still greater than its bulk value, is 
suppressed relative to that of the interior layers. 

In the full thin 
film model, the combined effects of terms (ii) and (iii) compete with 
the suppression of ferroelectricity by the dipolar contribution of 
term (i). The surface enhancement and characteristic shape of the 
polarization profile is preserved by the inclusion of term (i), while 
the magnitude of the overall enhancement is lowered, so that the 
values of polarization in the interior approach the bulk value.

When comparing to observations on ultrathin PbTiO$_3$ films, several factors must be 
taken into account. First, substrates will generally induce epitaxial stress, 
which can significantly change the ferroelectric state of the film 
\cite{pertsev}. The screening of the depolarization field by surface charges can 
be expected to be less effective than for the ideal case of perfectly conducting 
plates considered here, which would act to reduce the polarization enhancement
\cite{binder}. Finally, the finite conductivity of PbTiO3 could lead to the 
creation of a depletion layer and related effects \cite{watanabe} not included 
in this model. Nevertheless, the plausibility of surface polarization 
enhancement warrants further investigation.

In conclusion, we have constructed a microscopic ${\cal H}_{eff}$ for 
the study of PbTiO$_3$ thin films under stress-free and short-circuit 
boundary conditions.
This model shows that (001) PbTiO$_3$ films as thin as three unit 
cells exhibit a perpendicularly polarized ferroelectric ground state, 
with significant enhancement of the polarization at the surface. The 
ferroelectric instability is consistent with recent observations, 
suggesting further theoretical and experimental investigation.

{\bf Acknowledgements}

We thank Ch. Ahn, M. Buttiker, S. Desu, R. Ramesh, J. F. Scott, 
J.-M. Triscone and T. Tybell for stimulating discussions. We 
acknowledge 
the Aspen Center for Physics and the support of ONR Grant 
N00014-97-0047.

\newpage

\begin{table}
\caption{Energy per unit cell in eV (in reference to the paraelectric 
state) and macroscopic strain as a function of thickness ($N$) for 
perpendicularly ($\perp$) and in-plane ($\parallel$) polarized films.
The last column refers to the {\it simplified} model (see text).
For comparison, in the tetragonal bulk phase: $E=-0.086$ eV, 
$\epsilon_{xx}=\epsilon_{yy}=-0.015$ and $\epsilon_{zz}=0.064$. }
\label{Table.energy}
\begin{tabular}{lccccccr}
$N$   & $E_{tot}^{\parallel}$ 
&$\epsilon_{xx}^{\parallel}=\epsilon_{zz}^{\parallel}$
&$\epsilon_{yy}^{\parallel}$ & $E_{tot}^{\perp}$ 
&$\epsilon_{xx}^{\perp}=\epsilon_{yy}^{\perp}$
&$\epsilon_{zz}^{\perp}$ &$E_{model}^{\perp}$\\
\hline
3   &-0.110 &$-0.017$ &$0.071$   & $-0.320$ &$-0.023$ &$0.098$   
&$-0.309$ \\
5   &-0.076 &$-0.014$ &$0.061$   & $-0.217$ &$-0.020$ &$0.084$   
&$-0.210$ \\  
7   &-0.072 &$-0.014$ &$0.060$   & $-0.163$ &$-0.018$ &$0.076$ 
&$-0.172$ \\  
\end{tabular}
\end{table}

\begin{table}
\caption{Parameters of the simplified ${\cal H}_{eff}$ when energies 
are in eV, lengths in \AA $\;$ and $\xi_{n}$ is normalized to $a_{0}$.}
\label{Table.para}
\begin{tabular}{lcccccccccr}
$A_l$ & $-66.2664$  & &$a$        &$+118.046$   
& &$u_{conv}$    &$-14.3997$   & &$C_{11}$   &$+117.9$ \\
$B$ & $+6.906 \times 10^{3}$  & &$b$        &$+14.718$   
& &$u_{self}$  &$+32.7745 $   & &$C_{12}$   &$+51.5$\\
$C$ & $-1.658 \times 10^{5}$  & &$\Delta A_l$ &$-46.866$   
& &$u_{surf}$  &$-0.0472$   & &$g_{0}$    &$-107.7$\\  
$D$ & $+9.630 \times 10^{6}$  & &$\Delta a$ &$+26.598$   
& &$u_{nn}$    &$-2.3788$   & &$g_{1}$    &$-790.3$ \\  
\end{tabular}
\end{table}


\begin{references}
\bibitem{plates} 
For a fully quantitative assessment of the effects 
of electrical boundary conditions, it is necessary to move the plates at the 
level of the interionic Hamiltonian, before the projection. This 
would also lead to slight modifications of the short-range surface 
parameters.
\end{references}
\end{document}